\documentclass[11pt]{article}
\usepackage{natbib, graphicx, setspace, amsmath, amssymb, subfigure, url, multirow, booktabs, verbatim, bm,threeparttable,algorithm, color}
\usepackage{amsthm, rotating}
\usepackage{longtable}
\usepackage{float}

\allowdisplaybreaks[3]
\newtheorem{theorem}{Theorem}

\begin{document}

\setlength{\textheight}{575pt}
\setlength{\baselineskip}{23pt}

\title{\sc Pan-disease clustering analysis of the trend of period prevalence}

\author{Sneha Jadhav$^{1*}$, Chenjin Ma$^{2,1*}$, Yefei Jiang$^3$, Ben-Chang Shia$^4$, Shuangge Ma$^1$}

\maketitle

\begin{center}

{\it
$^1$Department of Biostatistics, Yale University\\
$^2$School of Statistics, Renmin University of China\\
$^3$Graduate Institute of Business Administration, College of Management, Fu Jen Catholic University\\
$^4$College of Management, Taipei Medical University\\
$^*$Joint first authors\\
For contact: stat1001@tmu.edu.tw (Shia) and shuangge.ma@yale.edu (Ma)}
\end{center}

\begin{abstract}
For all diseases, prevalence has been carefully studied. In the ``classic'' paradigm, the prevalence of different diseases has usually been studied separately. Accumulating evidences have shown that diseases can be ``correlated''. The joint analysis of prevalence of multiple diseases can provide important insights beyond individual-disease analysis, however, has not been well conducted. In this study, we take advantage of the uniquely valuable Taiwan National Health Insurance Research Database (NHIRD), and conduct a pan-disease analysis of period prevalence trend. The goal is to identify clusters within which diseases share similar period prevalence trends. For this purpose, a novel penalization pursuit approach is developed, which has an intuitive formulation and satisfactory properties. In data analysis, the period prevalence values are computed using records on close to 1 million subjects and 14 years of observation. For 405 diseases, 35 nontrivial clusters (with sizes larger than one) and 27 trivial clusters (with sizes one) are identified. The results differ significantly from those of the alternatives. A closer examination suggests that the clustering results have sound interpretations. This study is the first to conduct a pan-disease clustering analysis of prevalence trend using the uniquely valuable NHIRD data and can have important value in multiple aspects.
\end{abstract}


\section{Introduction}

For most if not all diseases, prevalence, both its specific values and temporal and spatial trends, have been well examined. The analysis of disease prevalence has important implications for the allocation and planning of medical resources, identification of risk factors, development of interventions, and other purposes. Accordingly, statistical methods have been extensively developed for analyzing disease prevalence, ranging from simple summary statistics to advanced nonparametric fitting. We refer to the literature for detailed discussions on clinical interpretation and application \citep{Rothman:2012} and statistical methodology \citep{Keiding:1991}.

Under the ``classic'' paradigm, prevalence, as well as other outcomes, have been studied for each disease separately. Accumulating evidences have shown that diseases can be ``correlated''. Examples include the positive correlations in the occurrence, prognosis, and other outcomes of breast cancer and ovarian cancer, which are attributable to the shared molecular risk factors and hormone-related treatment regimens. Another example is the positive correlations in the occurrence and other outcomes of asthma, lung cancer, and other respiratory diseases, which are attributable to the shared environmental and genetic risk factors. Recent studies have shown that the joint analysis of multiple diseases can lead to important insights not shared by its individual-disease-based counterpart \citep{Rzhetsky:2007, goh}.

For prevalence, the outcome of interest in this study, the joint analysis of multiple diseases has been conducted.
For example, a prospective observational study conducted in the U.S. focuses on the prevalence of chest pain and acute myocardial infarction (MI) and suggests that patients without chest pain on presentation represent a large segment of the MI population and are at an increased risk for delays in seeking medical attention \citep{Canto:2000}.
Another example is the cancer metastasis networks, which have been established based on clinical data \citep{Chen:2009}. Despite several interesting findings, our literature review suggests that the existing joint analyses of prevalence have the following limitations. Some studies focus on a small number of pre-selected diseases (for example, heart diseases only) and do not have a ``global'' perspective. The interconnections in prevalence among diseases are still largely unknown, and it is desirable to conduct ``unguided learning'' and identify new interconnections. Some studies conduct cross-sectional analysis. It has been recognized that temporal and spatial variations can have more important implications than static values. There are also studies with limitations in quantitative analysis by adopting (overly) simple statistical methods. In the single-disease analysis of prevalence (and other outcomes), it has been shown that sophisticated statistical methods, for example nonparametric modeling, have the power to produce important findings missed by simple statistics. Some other studies are based on limited data, for example, collected from a small number of hospitals and communities. Such data, although interesting to a certain extent, have a selection bias problem and cannot describe the more valuable population-level properties.

\begin{figure}[h]
  \centering
    \includegraphics[width=6cm, height=4.2cm]{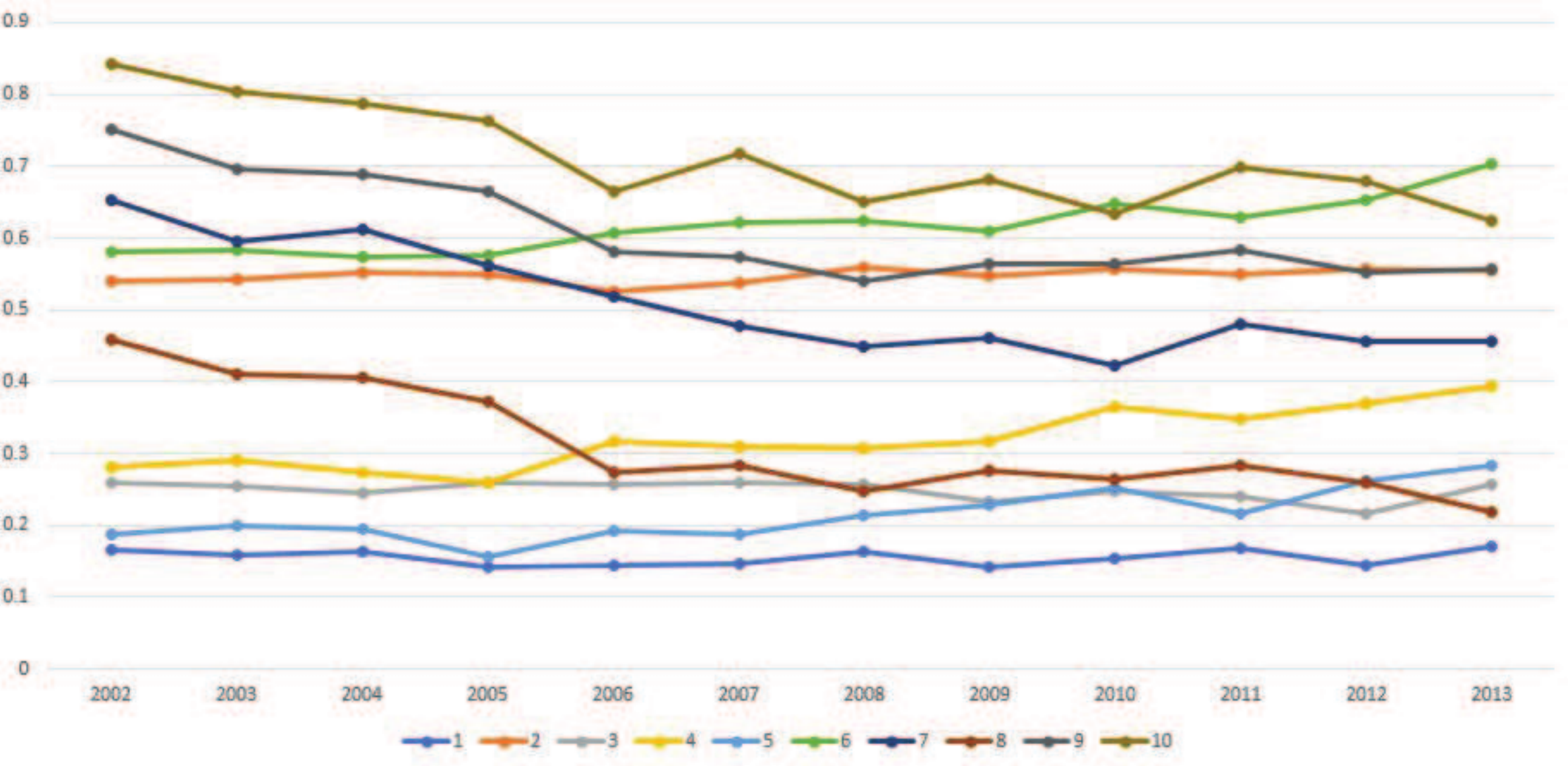}
    \includegraphics[width=5.8cm, height=4.2cm]{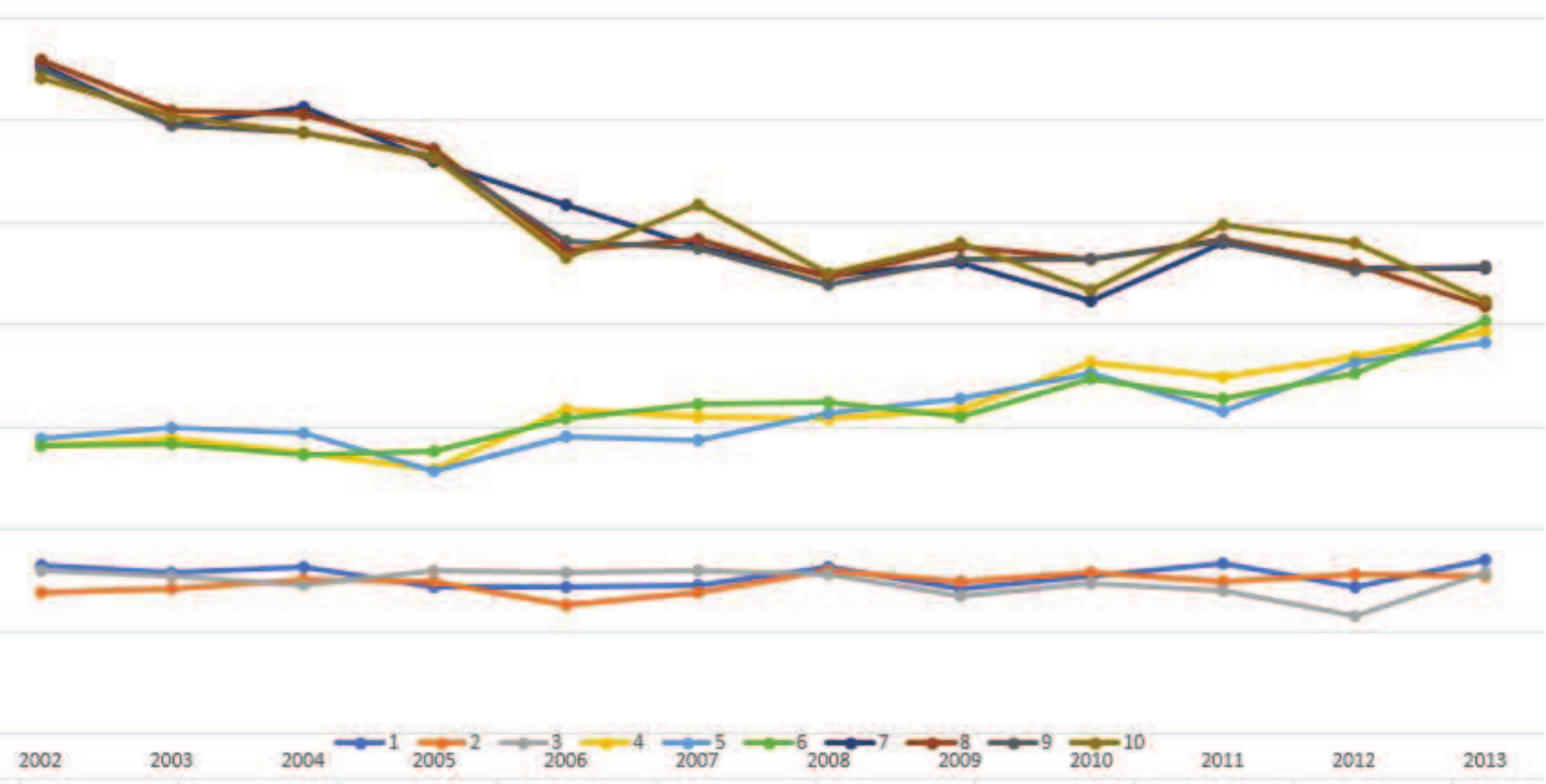}
  \vspace{1mm}{\small \caption{
 {\it Prevalence trends for ten diseases. Left: original measurements. Right: their clustering.}}}
  \label{fig:scheme}
\end{figure}

Our ultimate goal is to understand disease interconnections in terms of prevalence temporal trend. In this article, the specific goal is to identify clusters, within which diseases share similar prevalence temporal trends. To fix ideas, a symbolic presentation of our motivation and the proposed analysis is provided in Figure 1. In the left panel, we show the temporal trends of period prevalence for ten diseases, which may seem unstructured. However, if we ignore the absolute values and move the prevalence curves vertically, as shown in the right panel, the ten diseases clearly form three clusters, with those in the same cluster having similar trends.

In the literature, much attention has been paid to the magnitude of prevalence. In contrast, research on the shared temporal trends of diseases' prevalence has been extremely rare. However, it can have important implications beyond that on absolute magnitude. Specifically, the absolute magnitude of a disease' prevalence can be largely attributable to time-independent risk factors, such as genetic predisposition. In contrast, the variation of prevalence over time can be mostly attributable to time-varying factors, such as weather conditions, air/water quality, development of prevention programs, and others. If two diseases share similar temporal trends of prevalence, no matter their absolute magnitudes are close or not, it is sensible to conjecture that they share time-dependent risk factors and/or are affected by similar prevention programs. Such risk factors and prevention programs are more actionable than, for example, genetic predisposition. As such, the proposed analysis of shared temporal trends may have important practical implications not possessed by that on absolute magnitude.

This study can contribute beyond the existing literature in the following aspects. The proposed study objective, identification of diseases with shared prevalence trends via clustering analysis, innovatively differs from those in the existing studies, which usually focus on the static, absolute magnitudes. A large number of diseases are simultaneously analyzed. Avoiding pre-selection can lead to discoveries beyond existing knowledge.
The pan-disease analysis can facilitate a more ``global'' description of disease prevalence. The analyzed NHIRD data is uniquely valuable with a huge sample size and virtually no selection bias, which can lead to more credible and unbiased findings. In addition, a novel analysis method is implemented, which has a sound statistical basis and numerically outperforms multiple existing alternatives. It can have broad applications far beyond this study and thus independent value. Overall, this study may have significant biomedical as well as statistical merits.

\section{The National Health Insurance Research Database (NHIRD)}

Taiwan launched the single-payer National Health Insurance (NHI) program on March 1st, 1995. As of 2014, 99.9\% of Taiwan's population were enrolled. With the high cost of healthcare that is uninsured or covered by commercial insurance, almost all hospital/clinic-based healthcare has been going through NHI. Under NHI, hospitals and clinics are obligated to report detailed diagnosis (and treatment) data. Such data have been collected under the National Health Insurance Research Database (NHIRD) project (http://nhird.nhri.org.tw/en/). The NHI data have unique advantages, especially including unbiasedness (virtually the whole population is covered), comprehensiveness (information is available on all diagnosis/treatment episodes), and unity (all data collection follows the same protocol), and have served as the basis of a large number of epidemiological, clinical, health economic, and management studies (close to 400 publications on pubmed).

The proposed analysis may not be feasible with other databases. Specifically, hospital and community-based databases suffer from small sample sizes and a potential selection bias; Databases such as Medicare and Medicaid do not cover the general population and hence also have a selection bias problem; And those based on specific commercial insurance schemes are very difficult to access and suffer from a similar selection bias problem.

There are multiple ways of defining prevalence. In this study, we focus on period prevalence, which is defined as the percentage of the population with a certain disease within a fixed period of time (in our analysis, one year). The health care and insurance system in Taiwan is among the most developed, and undertreatment (presence of illness conditions that are not treated), although may exist, is expected to be rare. As such, the number of people treated with a certain disease (within one year) can provide a reasonable approximation to the number of people with this disease.

The analyzed data are extracted from NHIRD and contain records on 1 million randomly selected subjects (about 4.3\% of the Taiwan population) for the period of 2000-2013. Disease diagnosis (and hence period prevalence) information is extracted from both outpatient and inpatient treatments, using the NHIRD CD (Ambulatory care expenditures by visits) and DD (Inpatient expenditure by admissions) files. For disease definition, the ICD-9-CM code version 1992 (used before 2005) is transformed into the 2001 version. Records with the ICD-9-CM codes E and V (external causes of injury and supplemental classification), 630-679 (pregnancy, childbirth, and puerperium complications), and 760-999 (symptoms, sins, and ill-defined conditions) are removed from analysis following published literature. A small number of inconsistent records are removed. The final analyzed dataset contains records on 986,650 patients, and the disease period prevalence values are computed based on 173,355,725 outpatient and 1,381,749 inpatient treatment episodes. With the large number of observations, the accuracy of analysis may far exceed many peer studies.
ICD-9-CM contains more than 14,000 diseases, and directly using this code leads to very few diagnoses for many diseases. We apply the Phenome-Wide Association Study (PheWAS) vocabulary approach \citep{Wei:2017} and group the 14,000 ICD-9-CM diseases into 1,723 disease codes. It is realized that even with a huge sample size, the prevalence values for rare diseases may not be reliably estimated. As such, we remove diseases with extremely low prevalence. To make the analysis ``more interesting'', we further focus on two types of diseases. The first type of diseases are those with high prevalence and/or high mortality, for example upper respiratory tract infections, disease of the circulatory system, cancers, and others. Such diseases have significant public health implications. The second type of diseases are those with high clinical significance, for example, those with no effective treatment or clear causes, such as some cancers (e.g., melanoma), some diseases of the blood and blood-forming organs (e.g., autoimmune hemolytic anemias), congenital coagulation defects, and others.

A total of 405 diseases are included in the final analysis. This number is considerably larger than those in the existing studies. It is noted that the proposed analysis can be straightforwardly extended to all diseases. However, as for each disease the number of prevalence measurements (number of years) is rather limited, and as each disease is allowed to have its unique prevalence trend (details below), the analysis may become less reliable with a larger number of diseases.

Among the analyzed diseases, those with the highest prevalence include acute upper respiratory infections of multiple or unspecified sites, dental caries, acute bronchitis and bronchiolitis, acute sinusitis and others, all of which have been extensively studied in the literature. Those with the highest temporal variations include acute upper respiratory infections of multiple or unspecified sites, noninfectious gastroenteritis, gingivitis, essential hypertension, and others. Multiple different types of trends are observed, including virtually flat (for example, Takayasu's disease), increasing (for example, uterine leiomyoma), decreasing (for example, ulcerative colitis (chronic)), ``U'' shape (decreasing and then increasing; for example, benign neoplasm of thyroid glands), reversed U shape (for example, other diseases of the teeth and supporting structures), and others. More details are presented in Section 4 as well as available from the authors.

\section{Methods}

As can be partly seen from Figure 1, the proposed analysis basically poses a functional clustering (classification) problem. In the literature, multiple approaches have been developed. Examples include the functional K-means and those alike, which, under certain data/model assumptions, compute the $\ell_2$ distance between functions, and then conduct clustering in a similar way as for scalars. There are also methods that are more model-based. They assume that the observed functions (or their values at discrete time points) are generated from some  underlying parametric/nonparametric functions, which are then recovered for the purpose of clustering. For relevant discussions, we refer to Jacques \& Preda (2014), Abraham et~al.(2003), and Peng \& M{\"u}ller (2008).

\subsection{A penalization pursuit approach}

In ``simple'' clustering analysis with finite-dimensional random variables, it has been observed that no clustering approach is dominatingly better. It is thus always of interest to develop alternative methods. This can be especially true for functional clustering, considering the scarcity of available methods. In addition, as to be described below, the proposed method has multiple unique advantages.

In the analysis of NHIRD data, we have 14 period prevalence values (one per year) for each of the 405 diseases. Generically, assume that for each disease, the observed prevalence values are realizations of an unknown curve in $L^2[a,b]$, where $[a,b]$ is a finite interval. Denote $N$ as the number of diseases (sample size), $y_1(\cdot),...,y_N(\cdot)$ as the $N$ prevalence curves, and $t_1,...,t_T \in [a,b]$ as the $T$ time points of observation.
As described above, the proposed analysis focuses on trend as opposed to absolute magnitude. As such, prior to analysis, $y_i$'s are normalized to have ``means'' zero, and with a slight abuse of notations, we still use $y_i$'s to denote the normalized values.
Following a popular strategy in functional analysis, we conduct basis expansion. Specifically, consider
\begin{align}\label{model}
y_i(t_j)\approx \sum_{l=1}^{K} \beta_{il} \phi_l(t_j)+\epsilon_{ij},\,\epsilon_{ij}\overset{\mathrm{i.i.d}}{\sim} N(0,\sigma_i),\,i=1,...,N,\,j=1,...,T.
\end{align}
Here $\{\phi_l, l=1,\ldots, K\}$ is a set of $K$ basis functions in $L^2[a,b]$, and $\beta_{il}$'s are the unknown regression coefficients. In the literature, extensive works are available on choosing the form and number of basis functions and will not be reiterated here. Note that for the present problem, the assumption of independent errors can be sensible.

Denote $Y_i=(y_i(t_1), \ldots, y_i(t_T))^{\prime}$,
$B=\begin{bmatrix}
\phi_1({t_1})  & \dots & \phi_K({t_1}) \\
\vdots       &\ddots &  \vdots  \\
\phi_1({t_{T}})  & \dots & \phi_K({t_{T}})
\end{bmatrix}$, and
$\beta_i=(\beta_{i1},...,\beta_{iK})^{\prime}$. We consider the estimate
\begin{align}\label{obj1}
\{\hat{\beta}_j: j=1,\ldots, N\}= argmin\left\{ \sum_{i=1}^{N} \dfrac{1}{2} \|Y_i-B\beta_i\|^2+ \lambda \sum_{i< j} w_{ij}\| \beta_i-\beta_j \|\right\},
\end{align}
where $\lambda>0$ is a data-dependent tuning parameter and $w_{ij}$'s are data-dependent weights. In our numerical study, we set $w_{ij}=\|\hat{\beta}_i^{ols}-\hat{\beta}_j^{ols}\|^{-1}$ with
$\hat{\beta}_i^{ols}= argmin \|Y_i-B\beta_i\|^2$.
{\it Diseases $i$ and $j$ are concluded as in the same cluster if $\hat{\beta}_i=\hat{\beta}_j$.}

\noindent{\bf Rationale}
Here we adopt the popular basis expansion estimation approach. The proposed analysis can generate functional clusters as well as estimated prevalence curves for these clusters as a ``byproduct'', which may facilitate interpretation and other purposes. The proposed approach innovatively ``transforms'' clustering into a penalized pursuit problem, and relies on penalization to data-dependently determine the number of clusters, clustering structure, etc.
This strategy allows sufficient flexibility. For example, there is no assumption/constraint on the sizes of clusters, and it is possible that a disease forms a cluster of its own.
The strategy of clustering via penalization is pioneered in recent studies \citep{Shen:2010}. Significantly advancing from the literature, the proposed analysis is conducted on a large number of functions in the disease prevalence analysis. As each curve is represented by multiple basis functions and their regression coefficients, group penalization is needed. Here the group Lasso is adopted and can be replaced by other group penalties. An adaptive penalization is adopted to improve performance. There are extensive discussions in the literature on group and adaptive penalization and will not be reiterated here.
In ``classic'' nonparametric estimation, penalty is usually imposed on $\beta_i$'s. The present analysis goal differs from that in the existing studies, and the differences $\beta_i-\beta_j$'s are penalized to achieve clustering. When desirable, additional penalties can be imposed on $\beta_i$'s. In modeling, it is assumed that $\sigma_i$'s may vary across diseases. It is possible to estimate $\sigma_i$'s from OLS and construct a variance-weighted loss function. However, as a large number of variances need to be estimated, the unweighted objective function may be more stable and performs well as long as the variances are not too heterogeneous.

\subsection{Computation}

To simplify notation, consider the unweighted optimization
$$\text{min}_{\beta} \sum_{i=1}^{N} \frac{1}{2} \|y_i-B\beta_i\|^{2} +\lambda \sum_{i<j}\|\beta_i-\beta_j\|.$$
As in (\ref{obj1}) the weights are predetermined, its optimization can be conducted in exactly the same manner. The above optimization can be written as
$$min_{\beta} \sum_{i=1}^{N}  \frac{1}{2} \|y_i-B\beta_i\|^2 +\lambda \sum_{i<j}\|z_{ij}\|,\, \text{subject to }
\beta_i-\beta_j-z_{ij}=0.$$
Let $\beta=(\beta_1,...,\beta_N)',\,z=(z_{12},...,z_{(N-1)N})',\,v_{ij} \in R^{K},\, v=(v_{12},...,v_{(N-1)N})'.$
The augmented Lagrangian is given by:
$$ L_{\rho}(\beta,z,v)= \sum_{i=1}^{n} \frac{1}{2}\|y_i-B\beta_i\|^2 +\lambda \sum_{i<j}\|z_{ij}\|+\rho v_{ij}'(\beta_i-\beta_j-z_{ij})+\dfrac{\rho}{2}\| \beta_i-\beta_j-z_{ij}\|_2^{2}.  $$
For optimization, we adopt the ADMM technique, which is iterative with the following updates at iteration $m+1$:
\begin{align}
\beta_i^{(m+1)} &=arg min_{\beta_i} L_{\rho}(\beta,z^{(m)},v^{(m)}) ,\, i=1,...,N,\\
z_{ij}^{(m+1)} &=arg min_{z_{ij}} L_{\rho}(\beta^{(m)},z,v^{(m)})  ,\, i<j, \\
v_{ij}^{(m+1)}&=v_{ij}^{(m)}+\rho\sum_{i<j} (\beta_i^{(m+1)}-\beta_j^{(m+1)}-z_{ij}^{(m+1)}).
\end{align}
Taking derivative w.r.t $\beta_i$, we obtain
\begin{align*}
\dfrac{dL}{d\beta_i}&=-B'(y_i-B\beta_i)+\rho\sum_{i<j}v_{ij}
-\rho\sum_{i>j} v_{ji}+\rho\sum_{i<j}(\beta_i-\beta_j-z_{ij})-\rho\sum_{i>j}
(\beta_j-\beta_i-z_{ji}).
\end{align*}
With this equation, we can have
\begin{align*}
\beta_i^{(m+1)}=(B'B+(n-1)\rho)^{-1}\left(B'y_i+\rho\sum_{i<j}
v_{ij}^{(m)}-\rho\sum_{i>j} v_{ji}^{(m)}+\rho\left(\sum_{i<j}z_{ij}^{(m)}-\sum_{i>j}
z_{ji}^{(m)}-\sum_{i\neq j}\beta_{j}^{(m)}\right)\right).
\end{align*}
Taking derivative w.r.t $z_{ij}$ and equating to 0 yield
$$z_{ij}^{(m+1)}=\left(\|b_i^{(m+1)}-b_j^{(m+1)}+v_{ij}^{(m)}\|-
\dfrac{\lambda}{\rho}\right)_+\dfrac{b_i^{(m+1)}-b_j^{(m+1)}+
v_{ij}^{(m)}}{\|b_i^{(m+1)}-b_j^{(m+1)}+v_{ij}^{(m)}\|}.$$
Overall, with a fixed tuning parameter $\lambda$, the ADMM algorithm uses 0 or the OLS estimate as the starting value and iterates as described above until the difference between two consecutive estimates is small enough (such that convergence can be concluded).

\noindent\underline{Tuning parameter selection} $\lambda$ controls the degree of shrinkage and hence the number of clusters and clustering structure. In the extreme case with $\lambda=\infty$, all diseases are clustered together; and $\lambda=0$ leads to one disease per cluster.

It is noted that, different from most of the existing functional studies, the curves are not all iid realizations of the same underlying distribution. Potentially, a curve can be unique and form its own cluster. As such, commonly used methods such as the standard V-fold cross validation are not directly applicable. For each curve, we split the $T$ observations into a training and a testing set. The training set consists of the odd-numbered observations, and the testing set consists of the rest. Denote $\hat\beta_i^{train}$'s as the training set estimates. For curve $i$, denote $\hat{C}(i)$ as the estimated cluster membership. Assume that the first $\hat{C}_1$ clusters are nontrivial with sizes larger than one, and the rest $\hat{C}-\hat{C}_1$ clusters have sizes one. Denote $Y_i^{test}$ and $B^{test}$ as the testing set counterparts of $Y_i$ and $B$. We propose selecting $\lambda$ that minimizes
\begin{eqnarray}\label{cv}
\sum_{i=1}^N\|Y_i^{test}-B^{test}\hat\beta_i^{train}\|^2+
\sum_{c=1}^{\hat{C}_1} \sum_{i, j: \hat{C}(i)=\hat{C}(j)=c}
\|\hat\beta_i^{train}-\hat\beta_j^{train} \|+
\sum_{c=\hat{C}_1+1}^{\hat{C}} \|\hat\beta^{train}_{j:\hat{C}(j)=c} \|.
\end{eqnarray}
This has been partly motivated by the split-half tuning parameter selection for time series and other data, is expected to behave reasonably if the curves are relatively ``stable'', and  observed to have satisfactory numerical performance. In practice, we conduct a grid search for the optimal $\lambda$.

\subsection{Simulation}
Simulation is conducted to gain insights into the practical performance of the proposed method. In the first set of simulation, functions are generated as the cluster-specific mean functions plus random errors. Observations are then made for each function at 20 equally spaced time points with $t_1=0.1, \ldots, t_{20}=2$. For all $i$ and $j$, $\epsilon_{ij}\sim N(0, \sigma)$.

A total of eight clusters are defined as follows. \underline{Cluster 1}
Let $\psi_j(t),j=1,2$ be the monomial basis.
$y_i(t_j) = 2-\psi_1(t_j)-1.5\psi_2(t_j)+\epsilon_{ij}.$
\underline{Cluster 2} 	
Let $\psi_j(t),j=1,2$ be the monomial basis.
$y(t_j) = \psi_1(t_j)+1.5\psi_2(t_j)+\epsilon_{ij}.$
\underline{Cluster 3}
$y(t_j) = I(t_j\leq 1)(1-|t_j-1|)+ I(t_j> 1)+\epsilon_{ij}.$
\underline{Cluster 4}
$y(t_j) = 2I(t_j\leq 1)(|1-t_j|)+\epsilon_{ij}.$
\underline{Cluster 5}
Let $\psi_j(t),j=1,...5$ be the Fourier basis.
$y(t_j) = 3-2\psi_1(t_j)-0.8\psi_2(t_j)-\psi_3(t_j)-1.1\psi_4(t_j)-
0.5\psi_5(t_j)+\epsilon_{ij}.$
\underline{Cluster 6}
Let $\psi_j(t),j=1,...5$ be the Fourier basis.
$y(t_j) = 2\psi_1(t_j)+0.8\psi_2(t_j)+\psi_3(t_j)+1.1\psi_4(t_j)+
0.5\psi_5(t_j)+\epsilon_{ij}.$
\underline{Cluster 7}
$y(t_j) = 2(1-|t_j-1|)+\epsilon_{ij}.$
\underline{Cluster 8}
$y(t_j) = 2(1-|t_j|)+\epsilon_{ij}.$
To better comprehend the clusters, we present the mean functions (solid lines) as well as randomly realizations (dashed lines) in Figure 2.

\begin{figure}[h]
	\caption{Simulation: mean functions (solid lines) and random realizations (dashed lines).}
	\centering
	\includegraphics[height=9cm, width=16cm]{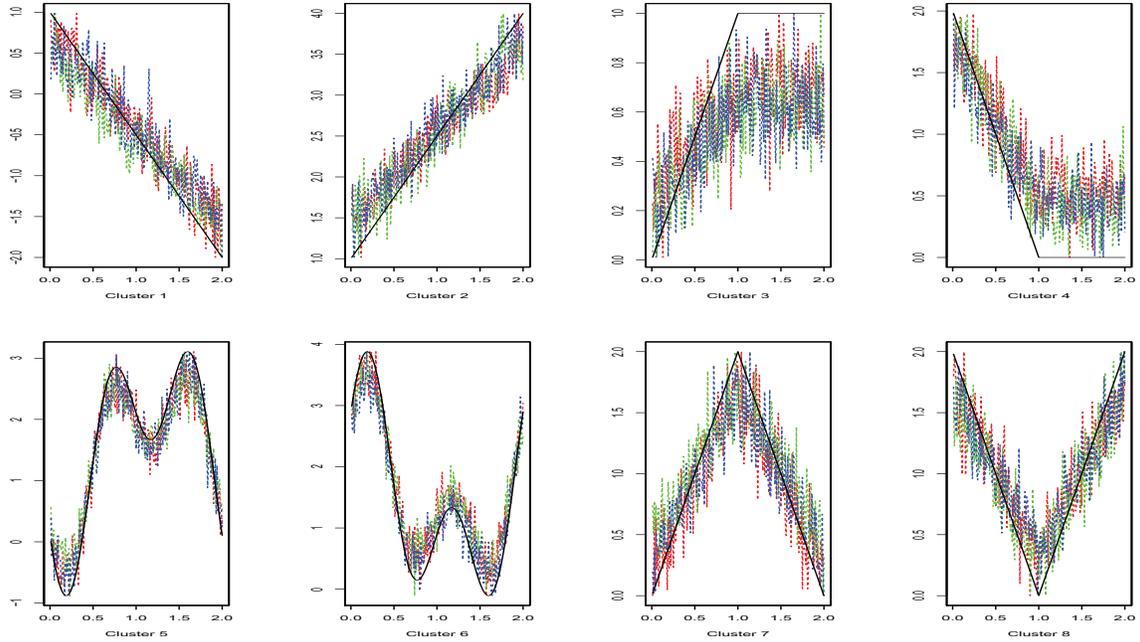}
	\label{Fig2}
\end{figure}

For the sizes of clusters, we consider two scenarios: Scenario 1 has equal sizes with $n_1=\ldots=n_8=12$, and Scenario 2 has unequal sizes with $n_1=11,\,n_2=17,\,n_3=15,\,n_4=9,\,n_5=11,\,n_6=13,\,n_7=14,\,n_8=10$.
In addition, we also consider different random error levels with $\sigma=0.1, 0.2, 0.3, 0.4$. In our analysis, clustering is of the highest interest. To evaluate clustering accuracy, consider the $N\times N$ matrix $\mathbf{A}$, whose $(i,j)$th element is 1 if curves $i$ and $j$ belong to the same cluster and 0 otherwise. Denote the estimate of $\mathbf{A}$ as $\hat{\mathbf{A}}$. We define clustering $Error=\sum_{i<j}|\mathbf{A}(i,j)-\hat{\mathbf{A}}(i,j)|$, which is the unnormalized error count.

\begin{table}[h]
	\def~{\hphantom{0}}
\centering
	\caption{Simulation I: mean clustering errors.}{%
		\begin{tabular}{ccccccccc}
\hline
 &	$\sigma$ & Proposed  &Alt.1   &Alt.2 & Alt.3 &Alt.4   &Alt.5 & Alt.6  \\
 \hline
Scenario 1 & 0.1	&	0.0	&	300.3	&	92.2	&	5.7	&	1344.8	&	0.0,	146.7	&	865.3	\\
           & 0.2 & 0.0	&	328.0	&	1019.0	&	34.6	&	1199.1	&	63.8,	110.4	&	889.9	\\
           & 0.3	&	9.0	&	455.2	&	2151.8	&	125.1	&	1277.2	&	144.3,	154.4	&	656.7	\\
           & 0.4	&	298.8	&	544.0	&	2381.3	&	177.0	&	1344.0	&	157.0,	165.1	&	458.5	\\
           \hline
Scenario 2& 0.1	&	0.5	&	259.0	&	121.4	&	3.4	&	1554.5	&	0.0,	125.2	&	580.1	\\
 & 0.2 &0.4	&	391.5	&	1021.0	&	44.8	&	1316.5	&	141.4,	166.6	&	739.0	\\
		&0.3	&	25.2	&	505.3	&	1958.3	&	145.0	&	1385.2	&	230.3,	227.7	&	567.6	\\
        &0.4	&	426.0	&	579.4	&	2167.6	&	207.7	&	1407.0	&	231.2,	236.9	&	638.4	\\
\hline
	\end{tabular}}
\end{table}

To better gauge performance of the proposed method, we also consider six relevant competitors.
[Alt.1] This method is proposed by Abraham et~al. (2003). It first fits data using B-splines (for each curve separately) and then conducts clustering of the estimated B-spline coefficients using the K-means approach. [Alt.2] The OLS estimates are first obtained for each curve separately. Then curves $i$ and $j$ are clustered together if $\|\hat{\beta}_i^{OLS}- \hat{\beta}_j^{OLS}\|< \eta$. Here $\eta$ serves as a tuning parameter and controls the number and structure of clustering. For determining the optimal $\eta$, we adopt an approach similar to that in Section 3.2. [Alt.3] This method is proposed by Bouveyron et~al. (2015) and based on mixture modeling. It takes functions as input. As such, we implement a smoothing method to obtain functions passing through the observed discretized points (more details can be found in Ramsay \&  Silverman (2005)). This method is implemented using the R package {\it funFEM}. Specifically, we specify the range of the number of clusters to be from 2 to 10 and select the ``all'' option, which allows for the implementation of all the 12 models proposed in this work and selects the optimal one.
[Alt.4] This is a curve clustering method using the functional random variable density approximation and proposed in Jacques \&  Preda (2013). It is realized using the R package {\it Funclustering}. We set the number of clusters as eight, which may lead to favorable performance.
[Alt.5] The FClust approach, which can be realized using the R package {\it fdapace}, conducts functional clustering and identification of data substructures for longitudinal and other functional data. In the original development, data-dependent determination of the number of clusters is not proposed. Here we consider two cases. The first is to set the number of clusters as the true. The second is to consider the number of clusters as true+1, which may serve as a sensitivity analysis.
[Alt.6] The funHDDC method, which can be realized using the R package {\it funHDDC}, conducts model-based clustering and identification of functional subspaces. It is originally developed for time series data.
For all seven methods, clustering errors are computed based on 100 replicates and presented in Table 1.

The proposed approach is observed to have competitive performance. For example under simulation Scenario 1 with $\sigma=0.3$, it has mean clustering error 9.0, compared to 455.2 (Alt.1), 2151.8 (Alt.2), 125.1(Alt.3), 1277.2 (Alt.4), 144.3 and 154.4 (Alt.5, with the number of clusters equal to true and true+1, respectively), and 656.7 (Alt.6). When the noise level increases, performance of all methods deteriorate, as expected. When $\sigma=0.4$ or higher, the proposed approach has reasonable but less competitive performance. A closer examination suggests that, for this specific simulation setting, this is largely caused by clusters 1 and 4. Figure 2 suggests that when the noise level is high enough, the two clusters become virtually indistinguishable. Alt.3 and Alt.5 are observed to have relatively ``robust'' performance when noise increases. It is noted that the simulated mean curves have ``nicer properties'' (compared to what may be encountered in general), which may favor Alt.3 taking the functional curves as input. It is also noted that, for Alt.5 and other alternatives, the favorable performance is attributable to the true number of clusters. Unlike the proposed, these methods do not have an easy way of data-dependently determining the cluster number. If the number of clusters is off, their performance may deteriorate significantly.

The second simulation is built on the first one. Here we set $\sigma=0.3$ and consider: Scenario 3 with $n_2=25, n_4=25, n_5=5$, $n_7=25$, and $n_1=n_3=n_6=n_8=0$, and Scenario 4 with $n_1=25, n_3=5, n_6=25, n_8=25$, and $n_2=n_4=n_5=n_7=0$.  Here there are four clusters, and the cluster sizes are highly unbalanced. Here we also examine performance of the proposed and alternative methods as $T$, the number of observation points, increases.

\begin{table}[h]
	\def~{\hphantom{0}}
\centering
	\caption{Simulation II: mean clustering errors.}{%
	\begin{tabular}{ccccccccc}
\hline
& $T$  & Proposed &Alt.1&Alt.2 & Alt.3 &Alt.4&Alt.5 & Alt.6  \\
\hline
Scenario 3 & 20 & 50.8	&	779.6	&	1361.7	&	572.7	&	911.6	&	0.0,	83.4	&	101.5	\\
		&40  &0.1	&	198.3	&	981.4	&	578.3	&	980.0	&	0.0,	77.3	&	733.1	\\
		&200 & 0.0	&	248.2	&	36.8	&	575.0	&	581.2	&	0.0,	80.8	&	496.0	\\
\hline
Scenario 4 & 20 &3.8	&	309.3	&	513.0	&	522.8	&	590.4	&	0.0,	88.7	&	186.9	\\
           &40  &0.5	&	224.4	&	27.6	&	575.2	&	622.3	&	0.0,	78.8	&	592.5	\\
  		   &200 & 0.0	&	316.6	&	0.4	&	584.8	&	502.3	&	0.0,	77.4	&	508.8	\\
  \hline
		\end{tabular}}
\end{table}

Table 2 suggests that as $T$ increases, performance of the proposed method improves fast and significantly. When $T=40$ or larger, the proposed approach virtually has no clustering error. Performance of Alt.2 also improves. This is reasonable, as loosely speaking, it shares a similar spirit as the proposed: both of them can be viewed as a certain form of thresholding. Performance of the other alternatives may be ``less sensitive'' to $T$.
A ``counterintuitive'' observation is that performance of some alternatives may get slightly worse when $T$ is large, which can be caused by the overly wiggly estimated functions. It is also observed that with Alt.5, if the number of clusters is set as true, performance can be extremely well. However, it becomes less satisfactory when the number of clusters is off.

\section{Analysis of NHIRD data}

Data described in Section 2 are analyzed using the proposed approach. A total of 35 nontrivial clusters (with sizes larger than one) are identified. In addition, there are also 27 diseases forming their own individual clusters. For the identified clusters, we plot the observed period prevalence curves (without normalization) in Figure 2. Here different colors are used to represent different prevalence levels. The detailed disease information is provided in Appendix II.

\begin{figure}[H]
\caption{Data analysis: clustering results using the proposed approach}
\centering
\includegraphics[height=17cm, width=14cm]{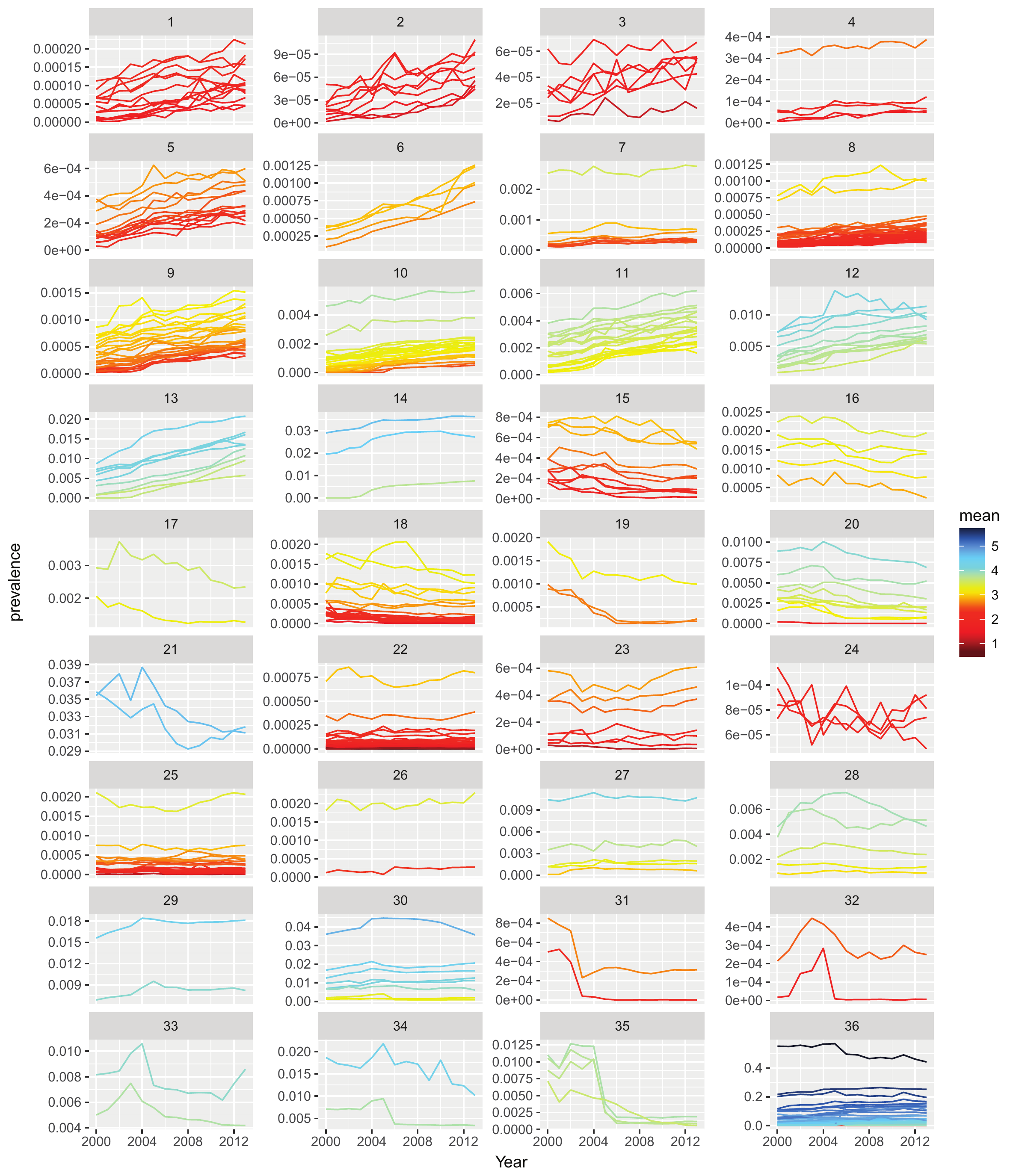}
\label{Fig3}
\end{figure}

A total of 201 diseases in the first 14 clusters in general have increasing trends. However, the shapes show considerable differences. As discussed in the literature, disease prevalence is affected by both genetic and non-genetic factors. With most if not all investigated diseases being genetically highly stable, the observed variations in prevalence are mostly attributable to non-genetic factors. For example, environmental factors, which usually vary over time, have been suggested as causal for the increase in prevalence of certain diseases. Factors identified in the literature include worse air quality, unhealthier dietary behaviors, stressful working conditions, and others. In addition, the development of more effective diagnostic tools may also contribute to the observed increasing prevalence \citep{Abdollah:2013}. A closer examination of Table 4 suggests that the clustering results may have sensible biomedical interpretations, with evidences reported in the literature observed worldwide as well as in Taiwan. Consider as an example cancers, which have important medical and public health implications and have been observed to have increasing trends. In Netherlands, Brazil, Iran, as well as many other countries, population-based studies have reported an increasing prevalence of melanoma. More related to the Taiwan population, a study conducted in mainland China analyzed the National Central Cancer Registry Database for the period of 1998 to 2007 and reported an increase in the prevalence of colorectal cancer. It is suggested that the increasing trend will continue in the near future \citep{Dai:2012}. The analysis of Taiwan Cancer Registry data, a database independent of NHIRD, suggests that oral cancer (OC) is one of the fastest growing malignancies in Taiwan, which is largely attributable to certain unhealthy lifestyle factors, including tobacco smoking, alcohol drinking, and betel quids chewing \citep{Chiang:2010}. Also included in clusters 1-14 and showing increasing prevalence include diseases such as Alzheimer's disease, hypertensive heart disease, chronic kidney disease, HIV infection, and others. They all have been observed to have increasing prevalence in the literature. The main goal of the proposed analysis is to identify clusters of diseases, which may be caused by shared risk factors. Examining clusters 1-14 suggests that there are sound interpretations for certain diseases being clustered together. For example, cluster 9 includes human papillomavirus (HPV) associated malignant neoplasm of uterus, cancer of tongue/nasopharynx, acute/prostatitis as well as orchitis and epididymitis. They share the same risk factor HPV \citep{Bonneau:2014}. In cluster 11, hyperglyceridemia and disorder of lipoid metabolism are confirmed risk factors of myocardial infarction and hypertension \citep{Austin:1998}. The latter can lead to myocardial infarction, hypertensive chronic kidney disease, and cerebrovascular disease \citep{Kario:2004}. In addition, acute renal failure is another risk factor of chronic kidney disease \citep{Lo:2009}. As core symptoms of depression \citep{Nutt:2008}, sleep disorders, depression, and major depressive disorder are grouped together (cluster 12). Our clustering analysis also identifies diseases sharing similar prevalence trends but do not yet have solid support from the literature. For example, acquired coagulation factor deficiency and renal osteodystrophy have very similar prevalence trends, but there is a lack of evidence for their connections. This can be caused by chance, connections with other diseases that are connected, as well as unidentified shared risk factors. More exploration is needed in the future.

In the next 7 clusters (15-21), a total of 55 diseases showing various decreasing prevalence trends. The decrease in disease prevalence can be caused by prevention interventions (lowering incidence), improvements in treatment and care, as well as disease-modifying interventions (preventing or slowing progression). Many of the decreasing trends observed in Figure 2 have also been reported in the literature for Taiwan as well as other countries/regions, and plausible causes have been suggested. For example, a decreasing trend of tuberculosis (TB) has been reported in most countries over the past decade, which is largely attributable to national tuberculosis control programs \citep{Ohmori:2002}. As above, many of the observed co-clusterings of diseases also have sound interpretations. For example, CNS infection/poliomyelitis and encephalitis are both included in cluster 15. CNS infection is a major risk factor of encephalitis \citep{Annegers:1988}. In cluster 18, acute rheumatic heart disease often leads to heart valve disorders and can cause ischemic heart disease as well as acute pulmonary heart disease when the condition deteriorates \citep{Young:2006}. As for the increasing trends, there are also new discoveries that warrant additional investigation. For example, cysts of oral soft tissues and duodenal ulcer have very similar trends in cluster 20. Their interconnection, if exists, has not been well established in the literature.

Diseases in cluster 22-30 have relatively flat prevalence with small fluctuations. Such diseases may be not heavily influenced by varying environmental risk factors, or the relevant environmental risk factors are rather ``constant'' over time, or the absolute values of prevalence are too low. Examples include obesity, umbilical hernia, congenital coagulation defects, and Takayasu's disease.

A total of 12 diseases with ``irregular'' trends are included in cluster 31-35. The ``irregularity'' in prevalence can be caused by the development of intervention programs, outbreaks of chronic diseases, and other factors. A close examination suggests that the observed irregular trends also have sound interpretations. Consider for example varicella infection. It has the highest age-specific incidence in children aged $<$9 years.  Overall, the prevalence of varicella shows a decreasing trend. More specifically, it fluctuates in the period of 2000 to 2004, and then decreases steadily from 2005 to 2013. The varicella vaccine was introduced in Taiwan in July 1997. Taipei City included the varicella vaccine in its free pediatric vaccination program for children aged $>$12 months in 1998. Taichung City/County followed this protocol in 1999. However, the varicella vaccination was not included as a part of routine immunization in all practices of child care until 2004 \citep{Chao:2012}. As such, fluctuations are observed for the 2000-2004 period; and a steady decrease is observed starting 2005. For the irregular trends, the interconnections among diseases can be more specific to Taiwan, and both intervention programs and disease outbreaks are highly disease specific.

There are also 27 diseases that form their own individual clusters. Most of them are common upper respiratory tract infections and oral diseases. The top eight with the highest prevalence include acute upper respiratory infections, acute bronchitis and bronchiolitis, acute sinusitis, gingivitis, periodontitis (acute or chronic), essential hypertension, noninfectious gastroenteritis, and acute pharyngitis, all with prevalence higher than 10 cases per 100 persons. The high prevalence values easily lead to high variations. As such, although some of these diseases may seem to have similar trends as those previously clustered, they do not belong to the same clusters. Examples include the acute upper respiratory infections (which is not clustered) and duodenal ulcer (cluster 20), both of which have seemingly similar decreasing trends but have different levels of variations and hence are not clustered together. Another example is gingivitis (un-clustered) and hypercholesterolemia (cluster 13).

\begin{table}[h]
	\def~{\hphantom{0}}
\centering
	\caption{Data analysis: discrepancy in clustering between different methods }{%
	\begin{tabular}{cccccccc}
\hline
& Proposed &Alt.1&Alt.2 & Alt.3 &Alt.4&Alt.5 & Alt.6  \\
\hline
Alt.1 & 0.17 & 0  \\
Alt.2 & 0.14 & 0.22 & 0  \\
Alt.3 & 0.82 & 0.69 & 0.75 & 0  \\
Alt.4 & 0.78 & 0.66 & 0.72 & 0.17 & 0  \\
Alt.5 & 0.13 & 0.19 & 0.19 & 0.72 & 0.68 & 0  \\
Alt.6 & 0.90 & 0.77 & 0.81 & 0.07 & 0.19 & 0.79 & 0\\
\hline
		\end{tabular}}
\end{table}

Clustering is also conducted using the six alternative approaches described in simulation. The resulted numbers of clusters are 20 (Alt.1), 19 (Alt.2), 10 (Alt.3), 12 (Alt.4), 10 (Alt.5), and 2 (Alt.6). The counterparts of Figure 3 for the alternatives are presented in Figures 4-9 in Appendix III. More detailed information on the clustered diseases is available from the authors. It is noted that the clustering error defined in simulation can also measure the discrepancy of two clustering results. The normalized clustering discrepancy values between any two of the seven approaches are presented in Table 3. It is observed that the results from the proposed clustering are relatively similar to those of Alt.1, Alt.2, and Alt.5, but quite different from the other three approaches. The six alternatives generate results from relatively similar (e.g., Alt.3 and Alt.6) to significantly different (e.g., Alt.2 and Alt.3). With practical data, it can be difficult to determine which clustering is more accurate. The better simulation performance and sound interpretations can provide some confidence to the proposed clustering.

\section{Discussion}

In the analysis of NHIRD and other data alike, the prevalence of various diseases is of great importance. Advancing from the literature, this study has taken a novel different strategy and focused on the similarity in diseases' prevalence trends under the functional clustering framework. Another highlight is that a pan-disease approach has been taken, under which a total of 405 diseases have been analyzed. This study is the first of its kind and, as described in Section 1, can have important practical implications. The results can be informative to the Taiwan population and health care/insurance system as well as other countries/regions and populations. A functional clustering method has been proposed using the innovative penalized pursuit technique. It has been shown to have competitive numerical performance and satisfactory statistical properties. It can have broad applications in functional data analysis far beyond this study.

This study can be potentially extended in multiple directions. Disease prevalence depends on multiple factors and varies across regions and populations. It can be of interest to conduct similar analysis for other regions/populations. It is also noted that this can be very challenging with the difficulty in obtaining data comparable to the NHIRD. The shared temporal trends of disease prevalence can be caused by multiple factors. Some examples of plausible interpretations have been provided. However, we defer to future research to more systematically and quantitatively quantify the causes of similar trends. It is also of interest to apply the proposed method to other functional data problems and conduct more comparisons.

\clearpage
\section*{Appendix I: Statistical properties}
In this study, we have been focused on methodological development and data analysis. To provide more insight into the proposed method, we also briefly examine its statistical properties.

For the convenience of notation, we rewrite the objective function as
\begin{align}\label{obj1}
\{\hat{\beta}_j: j=1,\ldots, N\}= argmin\left\{ \sum_{i=1}^{N} \dfrac{1}{2} \|Y_i-B\beta_i\|^2+ \lambda T K \sum_{i< j} w_{ij}\| \beta_i-\beta_j \|\right\}.
\end{align}
For this ``new'' objective function, we make the following assumptions:
\begin{itemize}
\item[(A1)]
$T\to \infty, K \to \infty ,\,\dfrac{K}{T} \to 0,\, \dfrac{K(\text{ log}(K))^3}{T} \to 0 $.

\item[(A2)]
$\lambda \sqrt{KT} \underset{ i,j}{\text{max }} \{w_{ij}| \beta_i\neq \beta_j \}= O_p(1)$.

\item[(A3)]
$\lambda \sqrt{TK} \underset{ i,j}{\text{min }} \{w_{ij}| \beta_i = \beta_j \}\to_P \infty$.
\end{itemize}
It is noted that for lucidity, we have made assumptions directly on the $w_{ij}$'s, which can be obtained from the estimation properties of the underlying functions. For example, under Assumption A1 and certain regularity conditions, Corollary 2.1 of \cite{he2000parameters} can lead to $\| \hat{\beta}^{ols}_i-\beta_i\|=O_p(\sqrt{K/T})$. In this case, $\lambda =O(1/\sqrt{TK})$ satisfies the assumptions.

\begin{theorem} [Estimation consistency]
$\| \hat{\beta}_i-\beta_i \|\to O_p(\sqrt{K/T}),\, \forall i$.
\end{theorem}

\noindent{\sc Proof.}
Let $U_i\in R^{K},\, u=(U_1,...,U_N)' \in R^{NK}$ and
$Q(\beta)=\sum_{i=1}^{N} \dfrac{1}{2} \|Y_i-B\beta_i\|^2+ \lambda TK \sum_{i<j\leq N} w_{ij}\| \beta_i-\beta_j \|$. We show that $$ \underset{T}{lim} inf P\left(\underset{u \in R^{NK}: \|U_i\|=\Delta}{inf} Q(\beta + U\sqrt{K}/\sqrt{T})>Q(\beta)  \right) >1- \epsilon.  $$
Consider
\begin{align*}
Q(\beta + U\sqrt{K}/\sqrt{T})-Q(\beta) &=\dfrac{K}{2T}\sum_{i=1}^{N} u'_i \left(B'B\right)u_i -\sum_{i=1}^{N}u_i'\left(\dfrac{\sqrt{K}}{\sqrt{T}}
B'(Y_i-B\beta_i)\right)\\
 & +\lambda TK\sum_{i<j\leq N}w_{ij}\|\beta_i-\beta_j-u_{ij}\sqrt{K}/\sqrt{T}\|-\sum_{i<j\leq N}w_{ij}\|\beta_i-\beta_j\| \\
 & =.5K\sum_{i=1}^{N} u'_i \left(\dfrac{B'B}{T}\right)u_i -\sum_{i=1}^{N}u_i'\left(\dfrac{\sqrt{K}}{\sqrt{T}}
 B'(Y_i-B\beta_i)\right)\\
 & +\lambda TK\sum_{i<j\leq N}w_{ij}\|\beta_i-\beta_j-u_{ij}\sqrt{K}/\sqrt{T}\|-\sum_{i<j:C(i)\neq C(j)}w_{ij}\|\beta_i-\beta_j\| \\
 & \geq .5K\sum_{i=1}^{N} u'_i \left(\dfrac{B'B}{T}\right)u_i -\sum_{i=1}^{N}u_i'\left(\dfrac{\sqrt{K}}{\sqrt{T}}
 B'(Y_i-B\beta_i)\right)\\
 & +\lambda KT\sum_{i<j:C(i)\neq C(j)}w_{ij}(\|\beta_{i}-\beta_{j}-u_{ij}
 \sqrt{K}/\sqrt{T}\|-\|\beta_{i}-\beta_{j}\|) \\
 & \geq.5K\sum_{i=1}^{N} u'_i \left(\dfrac{B'B}{T}\right)u_i -.5\sum_{i=1}^{N}u_i'\left(\dfrac{\sqrt{K}}
 {\sqrt{T}}B'(Y_i-B\beta_i)\right)\\
 & -\lambda\sqrt{KT}K\sum_{i<j:C(i)\neq C(j)}w_{ij}\|u_{ij}\|.
\end{align*}
Note that $\dfrac{B'B}{T} $ converges to the identity matrix. For each $i$, the first term is quadratic in $u_i$ and $O(K)$; the second term is linear in $u_i$ with its coefficient being $O_p(K)$; and finally from assumption A2, the last term is $O_p(K)$ and linear in $u_i$. Thus, with a properly chosen $\Delta$, the theorem can be proved.

\begin{theorem} (Clustering consistency)
Consider any $(i,j)$ with $C(i)=C(j)$. Then
$P(\hat{C}(i)=\hat{C}(j) )=P(\hat{\beta}_i=\hat{\beta}_j)\to 1$.
\end{theorem}

\noindent{\sc Proof.}
Consider a cluster including curves $y_1,y_2$, that is, $ \beta_1=\beta_{2}$. Below we show that
$P(\hat{\beta}_1\neq \hat{\beta}_{2})=0$ by contradiction.

Suppose that $\hat{\beta}_1\neq \hat{\beta}_{2}.$ Differentiating the objective function $Q(\beta)$ with respect to $\beta_1$ leads to the following normal equation:
\begin{eqnarray*}
&&B'(Y_1-B\hat{\beta}_1) + \dfrac{\lambda TK w_{12}}{\|\hat{\beta}_{1}-\hat{\beta}_{2}\|}(\hat{\beta}_{1}
-\hat{\beta}_{2})+ \sum_{j=3}^N\dfrac{\lambda TK w_{1j}}{\|\hat{\beta}_{1}-\hat{\beta}_{j}\|}(\hat{\beta}_{1}
-\hat{\beta}_{j}) =0,\\
&&B'(Y_1-B\beta_1) + B'B(\beta_1-\hat{\beta}_1)+\dfrac{\lambda TK w_{12}}{\|\hat{\beta}_{1}-\hat{\beta}_{2}\|}(\hat{\beta}_{1}
-\hat{\beta}_{2})+ \sum_{j=3}^N\dfrac{\lambda TK w_{1j}}{\|\hat{\beta}_{1}-\hat{\beta}_{j}\|}(\hat{\beta}_{1}
-\hat{\beta}_{j}) =0,\\
&&\dfrac{B'(Y_1-B\beta_1)}{\sqrt{TK}} + \dfrac{B'B}{T}\dfrac{\sqrt{T}}{\sqrt{K}}(\beta_1-\hat{\beta}_1)
+\dfrac{\lambda \sqrt{TK} w_{12}}{\|\hat{\beta}_{1}-\hat{\beta}_{2}\|}(\hat{\beta}_{1}-
\hat{\beta}_{2})+ \sum_{j=3}^N\dfrac{\lambda \sqrt{TK} w_{1j}}{\|\hat{\beta}_{1}-\hat{\beta}_{j}\|}(\hat{\beta}_{1}-
\hat{\beta}_{j}) =0.
\end{eqnarray*}
The first, second, and last terms are $O_p(1)$, and the third term dominates from Assumption A3, leading to a contradiction. Thus, $P(\hat{\beta_i} =\hat{\beta}_{2})=1 $.

\clearpage
\section*{Appendix II: Clustering results using the proposed approach}

	\begin{longtable}{cl}
     \caption{Diseases in each cluster(proposed method)}\\
\hline
& Diseases  \\
\hline
\endhead
\hline
\endfoot
Cluster 1 &	Allergic purpura \\
&	Angiodysplasia of intestine (without mention of hemorrhage)	\\
&	Bone cancer	\\
&	Cancer of bladder	\\
&	Cancer of other female genital organs, excluding uterus and ovary \\
&	Circadian rhythm sleep disorder	\\
&	Diaphragmatic hernia	\\
&	Disorders of phosphorus metabolism	\\
&	Exostosis of jaw	\\
&	Fluid overload	\\
&	Folate-deficiency anemia	\\
&	Glossodynia	\\
&	Melanomas of skin	\\
&	Ulceration of the lower GI tract \\
Cluster 2 &	Aneurysm of other specified artery	\\
&	Decreased libido	\\
&	Defibrination syndrome	\\
&	Disorders of iron metabolism	\\
&	Hyperchylomicronemia	\\
&	Incisional hernia	\\
&	Neoplasm of uncertain behavior of male genital organs	\\
&	Polycythemia vera	\\
&	Schizoid personality disorder	\\
Cluster 3 &	Acquired hemolytic anemias	\\
&	Autoimmune hemolytic anemias	\\
&	Benign neoplasm of parathyroid gland	\\
&	Carcinoma in situ of skin	\\
&	Dissociative disorder	\\
&	Lymphoid leukemia, chronic	\\
&	Other disorders of testis	\\
Cluster 4 &	Acquired coagulation factor deficiency	\\
&	Appendicitis	\\
&	Cancer of major salivary glands	\\
&	Renal osteodystrophy	\\
&	Reticulosarcoma	\\
Cluster 5 &	Antisocial/borderline personality disorder	\\
&	Aplastic anemia	\\
&	Benign neoplasm of male genital organs	\\
&	Cancer of liver and intrahepatic bile duct	\\
&	Dementia with cerebral degenerations	\\
&	Manlignant and unknown neoplasms of brain and nervous system	\\
&	Myoclonus	\\
&	Other disorders of lipoid metabolism	\\
&	Paranoid disorders	\\
&	Peripheral angiopathy in diseases classified elsewhere	\\
&	Pneumonia due to fungus (mycoses)	\\
&	Subarachnoid hemorrhage	\\
&	Subdural hemorrhage	\\
&	Tics and choreas	\\
Cluster 6 &	Alzheimer's disease	\\
&	Benign neoplasm of respiratory and intrathoracic organs	\\
&	Malignant neoplasm of bladder	\\
&	Nephritis and nephropathy in diseases classified elsewhere	\\
&	Thyroid cancer	\\
Cluster 7 &	Atherosclerosis of aorta	\\
&	Atherosclerosis of the extremities	\\
&	Other benign neoplasm of uterus	\\
&	Other deficiency anemia	\\
&	Other disorders of prostate	\\
&	Other inflammatory disorders of male genital organs	\\
&	Paralysis/spasm of vocal cords or larynx	\\
&	Proliferative glomerulonephritis	\\
Cluster 8 &	Abdominal aortic aneurysm	\\
&	Anomalies of jaw size/symmetry	\\
&	Benign neoplasm of other female genital organs	\\
&	Benign neoplasm of pituitary gland and craniopharyngeal duct (pouch)	\\
&	Cancer of brain	\\
&	Cancer of connective tissue	\\
&	Cancer of lip	\\
&	Cancer of oropharynx	\\
&	Cancer of other female genital organs	\\
&	Cancer of the gums	\\
&	Cerebral aneurysm	\\
&	Cerebral degeneration, unspecified	\\
&	Cyst of the salivary gland	\\
&	Degenerative disease of the spinal cord	\\
&	Disturbance of salivary secretion	\\
&	Fibroadenosis of breast	\\
&	Fibrosclerosis of breast	\\
&	Gastroesophageal laceration-hemorrhage syndrome	\\
&	Gram positive septicemia	\\
&	Hypersomnia	\\
&	Leukemia	\\
&	Lymphoid leukemia, acute	\\
&	Malignant neoplasm of gallbladder and extrahepatic bile ducts	\\
&	Malignant neoplasm of kidney, except pelvis	\\
&	Malignant neoplasm of small intestine, including duodenum	\\
&	Multiple myeloma	\\
&	Myeloid leukemia, acute	\\
&	Neoplasm of uncertain behavior of skin	\\
&	Nevus, non-neoplastic	\\
&	Nodular lymphoma	\\
&	Nonrheumatic pulmonary valve disorders	\\
&	Other disorders of arteries and arterioles	\\
&	Other disorders of carbohydrate transport and metabolism	\\
&	Other disorders of purine and pyrimidine metabolism	\\
&	Other hemoglobinopathies	\\
&	Other vitamin B12 deficiency anemia	\\
&	Posttraumatic stress disorder	\\
&	Pulmonary embolism and infarction, acute	\\
&	Raynaud's syndrome	\\
&	Spinocerebellar disease	\\
Cluster 9 &	Acidosis	\\
&	Acute posthemorrhagic anemia	\\
&	Acute prostatitis	\\
&	Benign neoplasm of brain, cranial nerves, meninges	\\
&	Benign neoplasm of kidney and other urinary organs	\\
&	Cancer of nasopharynx	\\
&	Cancer of tongue	\\
&	Cellulitis and abscess of trunk	\\
&	Cyst of kidney, acquired	\\
&	Delirium due to conditions classified elsewhere	\\
&	Disease of tricuspid valve	\\
&	Diseases of white blood cells	\\
&	Hemangioma and lymphangioma, any site	\\
&	Hydrocephalus	\\
&	Hypertensive heart and/or renal disease	\\
&	Malignant neoplasm of uterus	\\
&	Mitral valve stenosis and aortic valve stenosis	\\
&	Non-Hodgkins lymphoma	\\
&	Orchitis and epididymitis	\\
&	Other specified disorders of plasma protein metabolism	\\
&	Pancreatic cancer	\\
&	Parasomnia	\\
&	Prostatitis	\\
&	Pseudomonal pneumonia	\\
&	Secondary malignancy of brain/spine	\\
&	Secondary malignant neoplasm	\\
&	Sialoadenitis	\\
Cluster 10 &	Arthralgia/ankylosis of temporomandibular joint	\\
&	Benign neoplasm of lip, oral cavity, and pharynx	\\
&	Benign neoplasm of ovary	\\
&	Cancer of prostate	\\
&	Cellulitis and abscess of foot, toe	\\
&	Cerebral atherosclerosis	\\
&	Dementias	\\
&	Diseases of esophagus	\\
&	Disorders of the autonomic nervous system	\\
&	Essential tremor	\\
&	Extrapyramidal disease and abnormal movement disorders	\\
&	Glossitis	\\
&	HIV infection, symptomatic	\\
&	Hyperpotassemia	\\
&	Intracerebral hemorrhage	\\
&	Iron deficiency anemia secondary to blood loss (chronic)	\\
&	Malignant neoplasm of rectum, rectosigmoid junction, and anus	\\
&	Nasal polyps	\\
&	Nonrheumatic aortic valve disorders	\\
&	Other benign neoplasm of connective and other soft tissue	\\
&	Renal failure NOS	\\
&	Secondary malignancy of respiratory organs	\\
&	Secondary malignant neoplasm of liver	\\
&	Thrombocytopenia	\\
&	Ulcer of esophagus	\\
&	Vascular dementia	\\
Cluster 11 &	Acute renal failure	\\
&	Acute, but ill-defined cerebrovascular disease	\\
&	Agorophobia, social phobia, and panic disorder	\\
&	Bipolar	\\
&	Cervical intraepithelial neoplasia [CIN] [Cervical dysplasia]	\\
&	Chronic prostatitis	\\
&	Diseases of hard tissues of teeth	\\
&	Gastrojejunal ulcer	\\
&	Hepatitis NOS	\\
&	Hyperglyceridemia	\\
&	Hypertensive chronic kidney disease	\\
&	Hyposmolality and/or hyponatremia	\\
&	Iron deficiency anemias, unspecified or not due to blood loss	\\
&	Myocardial infarction	\\
&	Neoplasm of uncertain behavior	\\
&	Other persistent mental disorders due to conditions classified elsewhere	\\
&	Schizophrenia	\\
&	Temporomandibular joint disorder, unspecified	\\
&	Transient cerebral ischemia	\\
&	Unspecified disorder of lipoid metabolism	\\
&	Viral Enteritis	\\
Cluster 12 &	Angina pectoris	\\
&	Benign neoplasm of colon	\\
&	Bronchopneumonia and lung abscess	\\
&	Cellulitis and abscess of fingers/toes	\\
&	Depression	\\
&	Hemorrhage from gastrointestinal ulcer	\\
&	Herpes zoster	\\
&	Lump or mass in breast	\\
&	Major depressive disorder	\\
&	Nonrheumatic mitral valve disorders	\\
&	Septicemia	\\
&	Sleep disorders	\\
&	Viral hepatitis C	\\
Cluster 13 &	Chronic renal failure [CKD]	\\
&	Diseases of the larynx and vocal cords	\\
&	GERD	\\
&	Hypercholesterolemia	\\
&	Organic or persistent insomnia	\\
&	Senile dementia	\\
&	Uterine leiomyoma	\\
&	Viral hepatitis B	\\
&	Viral warts \& HPV	\\
Cluster 14 &	Cerebral artery occlusion, with cerebral infarction	\\
&	Hypertensive heart disease	\\
&	Stomatitis and mucositis (ulcerative)	\\
Cluster 15 &	Acute glomerulonephritis, NOS	\\
&	Aortic valve disease	\\
&	ASCVD	\\
&	Benign neoplasm of brain and other parts of nervous system	\\
&	Cervical cancer and dysplasia	\\
&	Encephalitis	\\
&	Hereditary disturbances in tooth structure	\\
&	Mitral valve disease	\\
&	Other CNS infection and poliomyelitis	\\
&	Peripheral vascular disease	\\
&	Vesicoureteral reflux	\\
Cluster 16 &	Arteritis NOS	\\
&	Atherosclerosis	\\
&	Inguinal hernia	\\
&	Nephritis and nephropathy with pathological lesion	\\
&	Viral pneumonia	\\
Cluster 17 &	Chronic tonsillitis and adenoiditis	\\
&	Diseases of pulp and periapical tissues	\\
Cluster 18 &	Acute appendicitis	\\
&	Acute febrile mucocutaneous lymph node syndrome (Kawasaki disease)	\\
&	Acute pulmonary heart disease	\\
&	Acute rheumatic heart disease	\\
&	Appendiceal conditions	\\
&	Atrophic gastritis	\\
&	Cerebral ischemia	\\
&	Diseases of the salivary glands	\\
&	Disorders of calcium/phosphorus metabolism	\\
&	Disorders of fluid, electrolyte, and acid-base balance	\\
&	Encephalitis, non-infectious	\\
&	Gout and other crystal arthropathies	\\
&	Heart valve disorders	\\
&	Inflammatory diseases of prostate	\\
&	Ischemic Heart Disease	\\
&	Lipoma of skin and subcutaneous tissue	\\
&	Neoplasm of uncertain behavior of breast	\\
&	Other hereditary hemolytic anemias	\\
&	Psychogenic and somatoform disorders	\\
&	Purpura and other hemorrhagic conditions	\\
&	Somatoform disorder	\\
&	Thrombotic microangiopathy	\\
Cluster 19 &	Benign neoplasm of unspecified sites	\\
&	Schizophrenia and other psychotic disorders	\\
&	Ulcerative colitis (chronic)	\\
Cluster 20 &	Acute and chronic tonsillitis	\\
&	Bacterial enteritis	\\
&	Benign neoplasm of uterus	\\
&	Cellulitis and abscess of oral soft tissues	\\
&	Cysts of oral soft tissues	\\
&	Disorders of lipoid metabolism	\\
&	Duodenal ulcer	\\
&	Loss of teeth or edentulism	\\
&	Tuberculosis	\\
&	Viral hepatitis	\\
Cluster 21 &	Chronic hepatitis	\\
&	Other local infections of skin and subcutaneous tissue	\\
Cluster 22 &	Acute vascular insufficiency of intestine	\\
&	Alkalosis	\\
&	Aneurysm and dissection of heart	\\
&	Aneurysm of iliac artery	\\
&	Anterior horn cell disease	\\
&	Azoospermia and oligospermia	\\
&	Benign neoplasm of eye	\\
&	Benign neoplasm of other endocrine glands and related structures	\\
&	Breast cancer [male]	\\
&	Cancer of bone and connective tissue	\\
&	Cancer of other endocrine glands	\\
&	Cancer of other lymphoid, histiocytic tissue	\\
&	Cancer within the respiratory system	\\
&	Celiac disease	\\
&	Colorectal cancer	\\
&	Congenital coagulation defects	\\
&	Dentofacial anomalies, including malocclusion	\\
&	Disease of capillaries	\\
&	Disorders of bilirubin excretion	\\
&	Disorders of magnesium metabolism	\\
&	Disorders of plasma protein metabolism	\\
&	Diverticulum of esophagus, acquired	\\
&	Hemoglobinuria	\\
&	Hemorrhagic disorder due to intrinsic circulating anticoagulants	\\
&	Hypersensitivity angiitis	\\
&	Intestinal infection due to protozoa	\\
&	Leprosy	\\
&	Lesions of stomach and duodenum	\\
&	Localized adiposity	\\
&	Malignant neoplasm of head, face, and neck	\\
&	Malignant neoplasm of other within the digestive organs and peritoneum	\\
&	Malignant neoplasm of unspecified male genital organ	\\
&	Megaloblastic anemia	\\
&	Moyamoya disease	\\
&	Nephritis and nephropathy without mention of glomerulonephritis	\\
&	Non-autoimmune hemolytic anemias	\\
&	Other specified diseases of the salivary glands	\\
&	Paraproteinemia	\\
&	Pernicious anemia	\\
&	Phenylketonuria [PKU]	\\
&	Pneumococcal pneumonia	\\
&	Polyarteritis nodosa	\\
&	Primary thrombocytopenia	\\
&	Qualitative platelet defects	\\
&	Renal sclerosis, NOS	\\
&	Sickle cell anemia	\\
&	Sideroblastic anemia	\\
&	Small kidney	\\
&	Specific nonpsychotic mental disorders due to brain damage	\\
&	Spermatocele	\\
&	Streptococcus infection	\\
&	Stricture and stenosis of esophagus	\\
&	Stricture of artery	\\
&	Takayasu¡¯s disease	\\
&	Thromboangiitis obliterans	\\
&	Vascular disorders of kidney/hypertrophy	\\
&	Vascular hamartomas and non-neoplastic nevi	\\
&	Von willebrand's disease	\\
Cluster 23 &	Arterial embolism and thrombosis	\\
&	Benign neoplasm of adrenal gland	\\
&	Benign neoplasm of bone and articular cartilage	\\
&	Benign neoplasm of thyroid glands	\\
&	Chondrocalcinosis	\\
&	Diseases of the jaws	\\
&	Intestinal e.coli	\\
Cluster 24 &	Infectious mononucleosis	\\
&	Malignant neoplasm of ovary and other uterine adnexa	\\
&	Multiple sclerosis	\\
&	Primary pulmonary hypertension	\\
Cluster 25 &	Acid-base balance disorder	\\
&	Amyloidosis	\\
&	Cancer of larynx, pharynx, nasal cavities	\\
&	Complications of gastrostomy, colostomy and enterostomy	\\
&	Disorders of penis	\\
&	Intestinal malabsorption (non-celiac)	\\
&	Lipoma	\\
&	Malignant neoplasm, other	\\
&	Mood disorders	\\
&	Non-proliferative glomerulonephritis	\\
&	Other cerebral degenerations	\\
&	Other specified benign mammary dysplasias	\\
&	Peripheral autonomic neuropathy	\\
&	Personality disorders	\\
&	Sexual and gender identity disorders	\\
&	Tooth complications likely association with other diseases	\\
&	Transient mental disorders due to conditions classified elsewhere	\\
&	Umbilical hernia	\\
&	Ventral hernia	\\
Cluster 26 &	Cystic mastopathy	\\
&	Overweight, obesity and other hyperalimentation	\\
Cluster 27 &	Bacterial pneumonia	\\
&	Gingival and periodontal diseases	\\
&	Nephrotic syndrome without mention of glomerulonephritis	\\
&	Obesity	\\
&	Ulceration of intestine	\\
Cluster 28 &	Bacterial infection NOS	\\
&	Balanoposthitis	\\
&	Chronic obstructive asthma	\\
&	Diseases of lips	\\
&	Other acute and subacute forms of ischemic heart disease	\\
Cluster 29 &	Chronic airway obstruction	\\
&	Other chronic ischemic heart disease, unspecified	\\
Cluster 30 &	Benign neoplasm of skin	\\
&	Chronic obstructive asthma with exacerbation	\\
&	Dysthymic disorder	\\
&	Erectile dysfunction [ED]	\\
&	Gout	\\
&	Gouty arthropathy	\\
&	Other diseases of the teeth and supporting structures	\\
&	Regional enteritis	\\
Cluster 31 &	Benign mammary dysplasias	\\
&	Chronic pulmonary heart disease	\\
Cluster 32 &	Coagulation defects	\\
&	Disorders of esophageal motility	\\
Cluster 33 &	Occlusion of cerebral arteries	\\
&	Periapical abscess	\\
Cluster 34 &	Cerebrovascular disease	\\
&	Viral infection	\\
Cluster 35 &	Disorders of function of stomach	\\
&	Gastritis and duodenitis	\\
&	Peptic ulcer (excl. esophageal)	\\
&	Varicella infection	\\
Cluster 36 &	Acute bronchitis and bronchiolitis	\\
&	Acute gastritis	\\
&	Acute pharyngitis	\\
&	Acute sinusitis	\\
&	Acute upper respiratory infections of multiple or unspecified sites	\\
&	Anemia of chronic disease	\\
&	Asthma	\\
&	Asthma with exacerbation	\\
&	Cancer, suspected or other	\\
&	Cellulitis and abscess of arm/hand	\\
&	Dyspepsia and other specified disorders of function of stomach	\\
&	Essential hypertension	\\
&	Gastritis and duodenitis, NOS	\\
&	Gingivitis	\\
&	Mixed hyperlipidemia	\\
&	Noninfectious gastroenteritis	\\
&	Oral aphthae	\\
&	Other anemias	\\
&	Other upper respiratory disease	\\
&	Periodontitis (acute or chronic)	\\
&	Reflux esophagitis	\\
&	Superficial cellulitis and abscess	\\
&	Dental caries	\\
&	Atopic/contact dermatitis due to other or unspecified	\\
&	Functional digestive disorders	\\
&	Type 2 diabetes	\\
&	Influenza	\\
  \hline
		\end{longtable}

\clearpage
\section*{Appendix III: Clustering results using the alternatives}
\begin{figure}[h]
	\caption{Clustering results using Alt.1.}
	\centering
	\includegraphics[height=9cm, width=16cm]{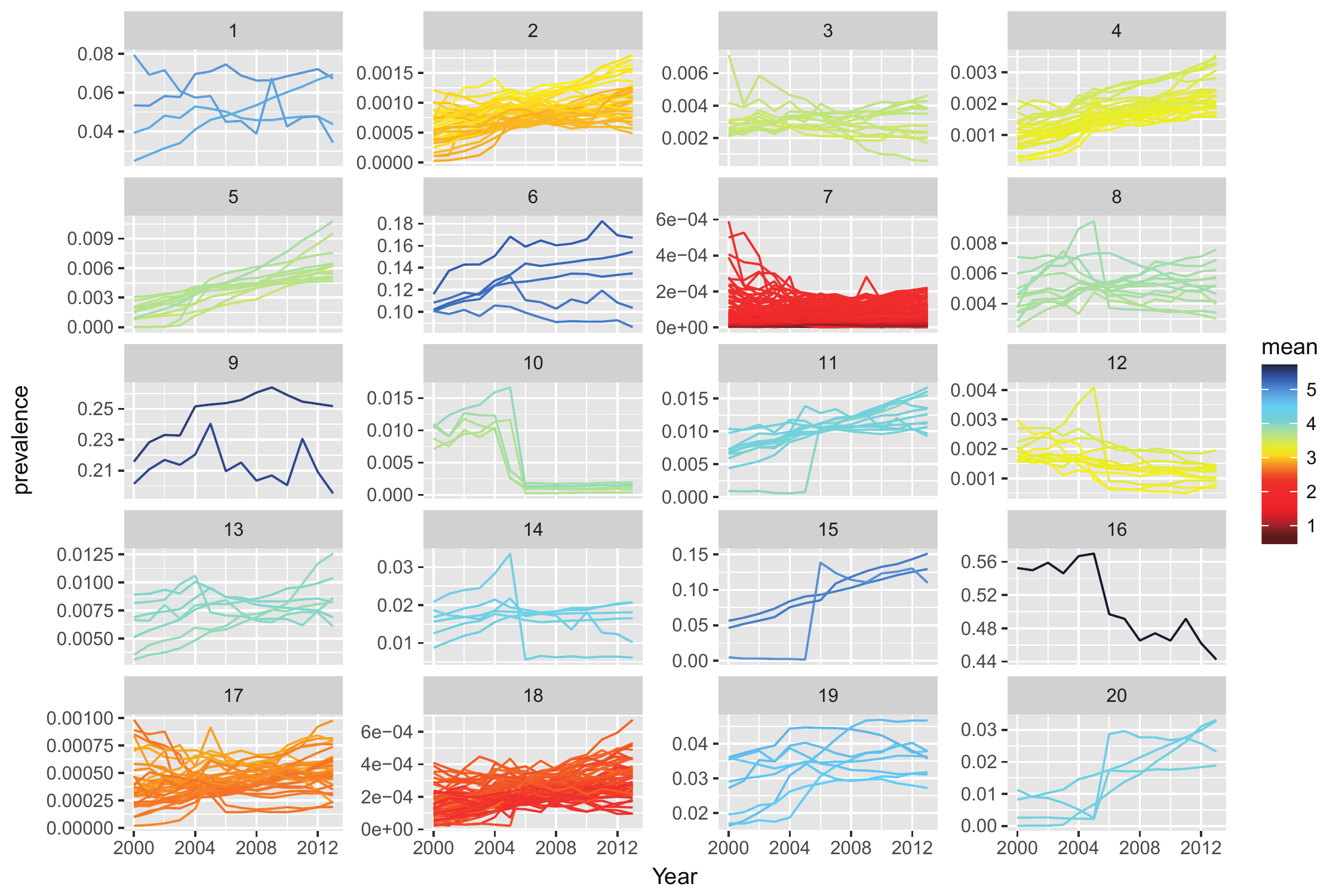}
	\label{Fig4}
\end{figure}
\begin{figure}[h]
	\caption{Clustering results using Alt.2.}
	\centering
	\includegraphics[height=9cm, width=16cm]{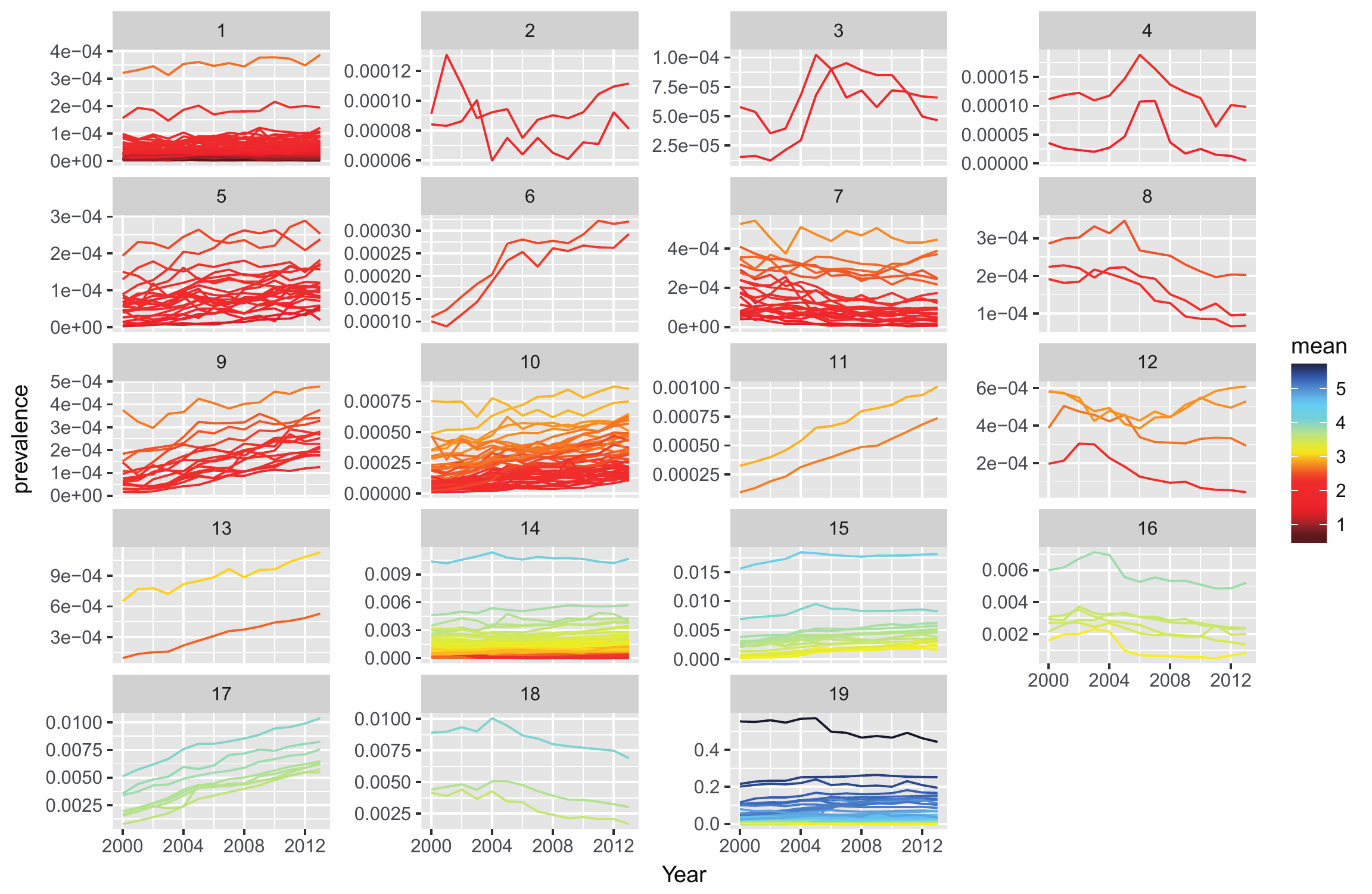}
	\label{Fig5}
\end{figure}

\begin{figure}[h]
	\caption{Clustering results using Alt.3.}
	\centering
	\includegraphics[height=8cm, width=16cm]{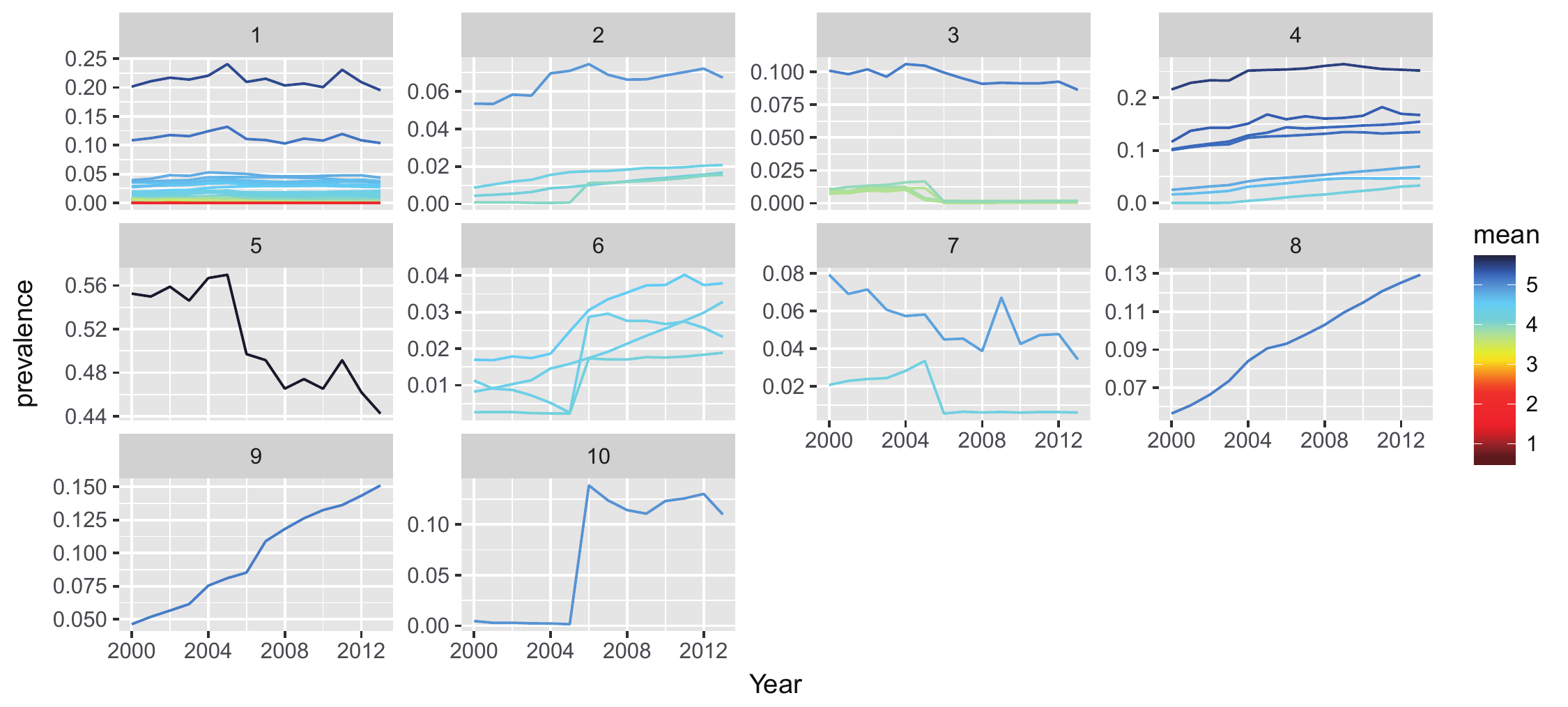}
	\label{Fig6}
\end{figure}
\begin{figure}[h]
	\caption{Clustering results using Alt.4.}
	\centering
	\includegraphics[height=8cm, width=16cm]{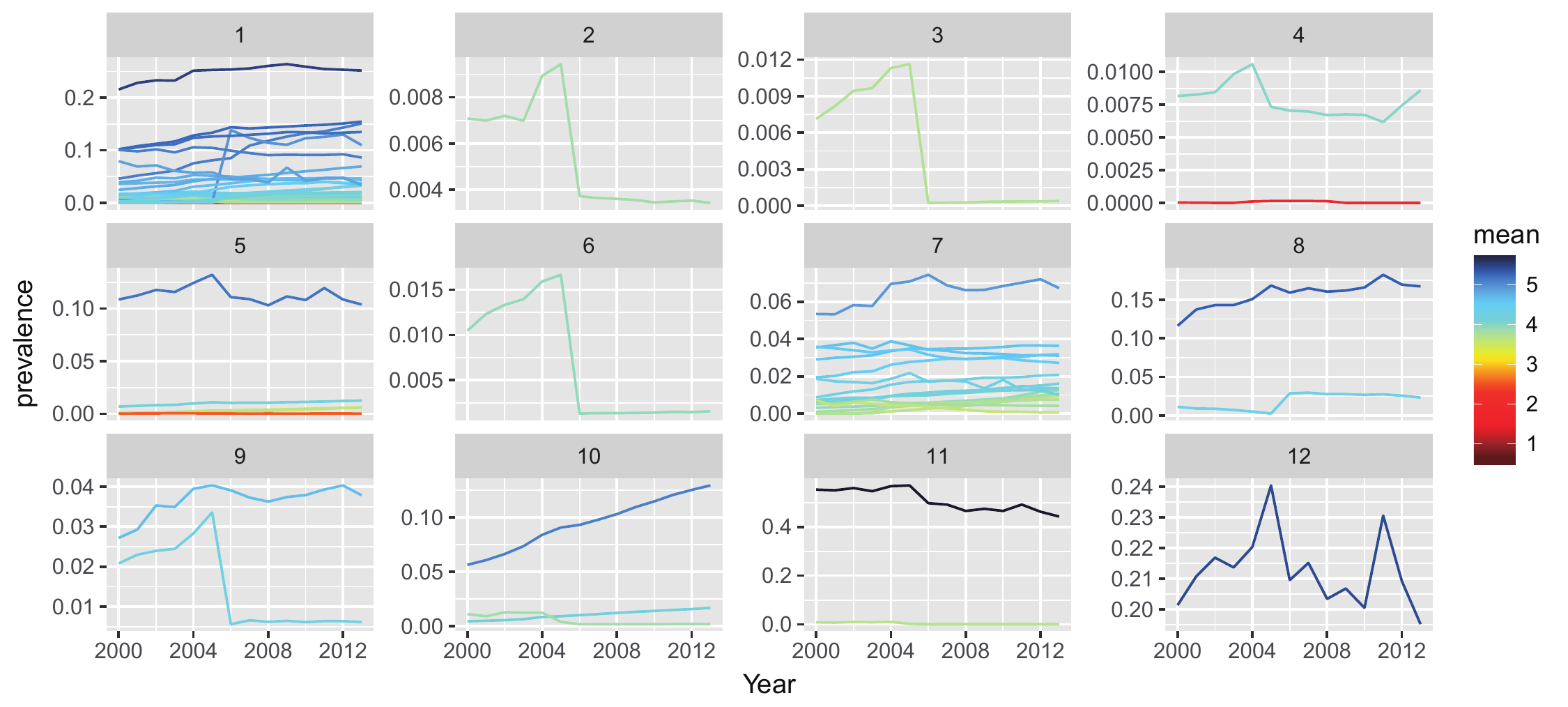}
	\label{Fig7}
\end{figure}
\begin{figure}[h]
	\caption{Clustering results using Alt.5.}
	\centering
	\includegraphics[height=10cm, width=16cm]{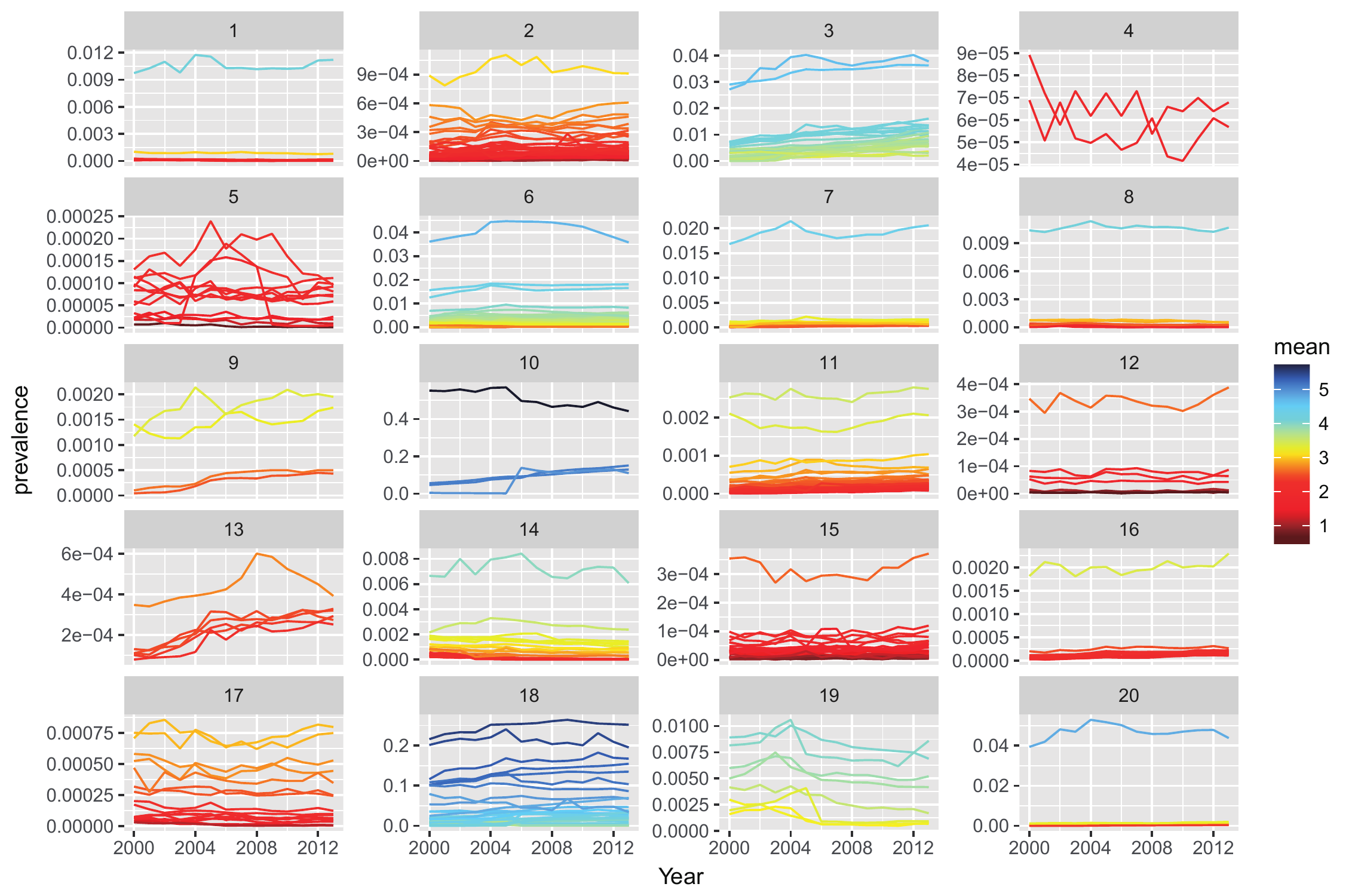}
	\label{Fig8}
\end{figure}
\begin{figure}[h]
	\caption{Clustering results using Alt.6.}
	\centering
	\includegraphics[height=7cm, width=16cm]{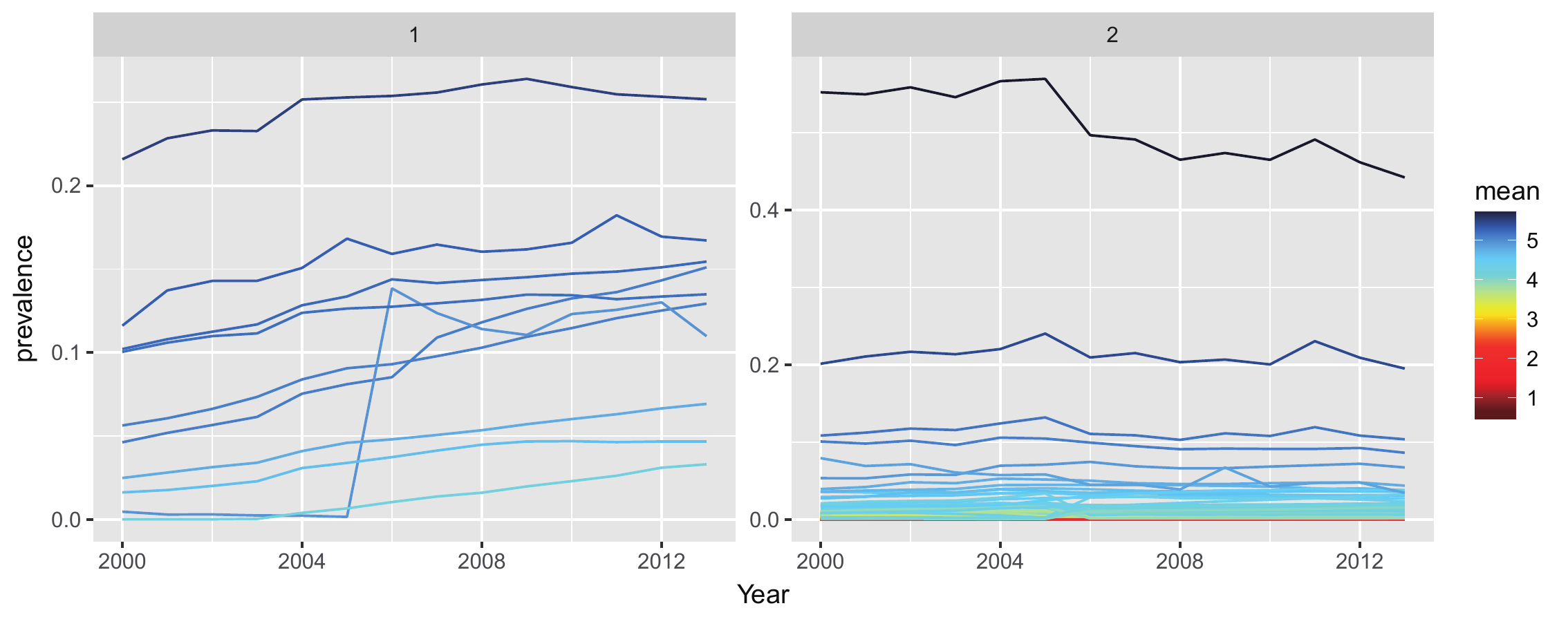}
	\label{Fig7}
\end{figure}

\end{document}